# Strong interlayer coupling and chiral flat-band cascades in twisted bilayer gratings

Daegwang Choi[1,†], Soon-Jae Lee[1,†], Seung-Woon Cho[1], Chan Bin Bark[2], Seik Pak[2], Dong-Jin Shin[1], Moon Jip Park[2,*] and Su-Hyun Gong[1,*]

[1]Department of Physics, Korea University, Seoul, 02841, Republic of Korea

[2]Department of Physics, Hanyang University, Seoul, 133-791, Republic of Korea

**Abstract**

From atomic crystals to macroscopic material structures, twisted bilayer systems have emerged as a promising route to control wave phenomena. In few-layer van der Waals (vdW) materials, however, the intrinsically weak interlayer coupling typically demands fine control of small twist angles to reach magic-angle conditions. Here, we show that one dimensional photonic crystal bilayers can overcome this limitation by accessing a regime of strong interlayer coupling—comparable to intralayer coupling. This strong coupling enables flat-band formation over a broad angular range even at large twist angles. We experimentally realize this regime by stacking WS$_2$ gratings using a two-step lithography method, resulting in ultra-wide chiral flat-band cascades in magic-angle twisted bilayer gratings. Our work not only provides a platform for designing photonic applications with a tunable localization but also explores a new regime of physics that unattainable in conventional solid-state based moiré systems.

**Main**

Twisted multilayer systems offer a promising platform for engineering wave mechanics beyond conventional lattice length scales[1,2]. Geometrically, for a given twist angle $\theta_t$ between two layers, and the lattice distance $a$, the superlattice period is amplified as $L_m \sim a/\theta_t$ leading to a significant enhancement of length scales[3]. A particularly intriguing phenomenon in such systems is the emergence of flat bands[4] and associated wave localization, which have played a key role in the discovery of superconductivity[5] and correlated insulators[6].

While atomic crystal structures and their vdW interactions are inherently dictated by material properties, photonic moiré superlattices offer unique advantages, such as the ability to precisely control interlayer interactions and engineer a diverse range of photonic crystal geometries[7]. The realizations of these effects in photonic systems are appealing for potential applications such as slow light and strong light-matter coupling[8,9]. In this context, moiré physics has been explored using various two dimensional (2D) lattice configurations, including Bravais, honeycomb, and Kagome lattices[10–19]. Twisted one dimensional (1D) gratings have also been reported[20–25], primarily focusing on optical chirality, while flat-band phenomena have remained barely explored in these systems. Moreover, most photonic moiré flat bands reported so far have been observed in regions of very small momentum[12]. The moiré superlattice size becomes extremely large at small twist angles, limiting the accessible momentum space where flat bands appear.

In this work, we provide experimental demonstrations of moiré systems exhibiting strong interlayer coupling and cascades of ultra-wide chiral flat bands. The strong interlayer coupling is achieved in the twisted bilayer grating (TBG) structure. Correspondingly, the resonant scattering through zone folding reveals the continuous inter-dimensional crossover from 1D grating ($\theta_t = 0°$) to 2D square lattice ($\theta_t = 90°$). The resonant conditions enabled by the strong coupling produces the robust magic-angle sequences for flat bands, with the geometrical predictable interpretation. These flat bands appear sequentially with varying twist angles, forming cascades of flat-band states. Thus, the distinct symmetry of twisted bilayer 1D grating structures leads to unconventional scattering mechanisms, setting them apart from their 2D counterparts.

To realize this concept experimentally, we implement TBG structures using WS$_2$, a transition metal dichalcogenide (TMD) which enables controllable layer stacking without the constraint of lattice matching, while also minimizing air gaps between rigid layers, making it highly suitable for constructing well-defined bilayer photonic structures[26]. In such bilayer structure, the breaking of vertical symmetry induces optical chirality leading to the emergence of chiral flat bands in the TBG structure. Notably, these flat bands exhibit ultra-wide characteristics compared to previously reported chiral flat bands in photonic systems[27,28].

**Twisted bilayer grating structures**

We consider a TBG structure, as schematically illustrated in Fig.1a. This structure consists of two identical dielectric WS$_2$ gratings, stacked with a twist angle $\theta_t$. Unlike 2D lattices, TBG exhibits distinct geometric characteristics (Fig. 1b), as they maintain a well-defined translational symmetry at arbitrary twist angles. These lattice transformations belong to the unimodular group, capturing smooth deformations such as squeezing, shear stress, and rotations. Figure 1b shows the top and side view of the unit cell of the TBG structure for numerical calculations. The top and bottom WS$_2$ gratings have identical parameters: thickness $h$, grating period $a$, and grating width $w$, and are placed on a quartz substrate.

We analyze the photonic dispersion of the TBG structures using rigorous coupled-wave analysis (RCWA). Figure 1c shows the simulated angle-revolved transmittance spectra along the *x*-axis with varying twist angle. At small twist angles, the spectrum reveals a Dirac-like band, characteristic of 1D gratings[29]. As $\theta_t$ increases, additional quadratic dispersions emerge from the low-energy region and gradually blue-shifts, while the Dirac-like band remains nearly unchanged.

The appearance of multiple parabolic branches and the evolution of their energies suggest that they originate from zone folding induced by the moiré superlattice in the TBG structures. Over a broad range of twist angles, clear avoided crossings between the Dirac-like and quadratic bands are observed, indicating strong interband coupling. As a result of this strong hybridization, a flat band emerges at a twist angle of approximately $\theta_t = 44.0°$. Notably, the flat band persists over an exceptionally wide angular range of about ±30°.

As the twist angle increases further, an additional flat band emerges near $\theta_t = 64.0°$, following similar band evolution. The detailed evolution of band dispersions with varying twist angle is provided in Extended Data Fig.1. We also confirm the presence of flat bands in TBG structures with varying grating parameters, including thickness and grating period. While the energy level of the flat band shifts with these changes, the twist angles at which flat bands form—the so-called magic angles—remain nearly invariant. This robustness highlights the geometric origin of flat-band formation in these structures (see Extended Data Figs. 2 and 3).

Figures 1d and 1e present the simulated electric-displacement distributions for $\theta_t = 44.0°$ and $64.0°$, respectively. For each case, the left and middle panels show the *E*-field profiles in the *x-y* plane taken at the center of the top and bottom WS$_2$ layers. The right panels display the corresponding *x-z* cross-sectional distribution, obtained by slicing horizontally along the central line of the *x-y* plane (red dashed lines). The white scale bars represent 200 nm. In both cases, the electric fields are well confined within the TBG region, confirming that the observed photonic bands originate from guided mode resonance (GMR) modes. Moreover, at the flat-band energies, the electric field exhibits pronounced spatial localization within the overlapping grating regions. This strong confinement is attributed to the near-zero group velocity at the flat-band condition.

**Mechanism of ultra-wide flat-band cascades**

The avoided level crossing originates from Umklapp scattering that couples the states at $k$ and $k'$ across the extended Brillouin zone. To elucidate the resulting flat-band cascades, we employ a coupled-mode Bloch Hamiltonian and analyze the bands in the extended Brillouin zone (BZ) framework, where the coupled momenta $k$ and $k'$ satisfy the relation as,

$$k' = k + m\vec{b}_1 + n\vec{b}_2, (m, n \in \mathbb{Z}). \qquad (1)$$

Here $\vec{b}_1, \vec{b}_2$ are the primitive reciprocal lattice vectors. Fig.2a illustrates reciprocal space in the extended BZ, where the dots represent momenta coupled through the zone folding for twist angles $\theta_t$= 0°, 46.5°, and 63.0°. In the absence of interlayer interaction, only between the lateral momenta along the 1D directions (*m* = 0 or *n* = 0, indicated by red dots) are coupled by the lowest-order scattering, thereby preserving the inherent character of the single-layer structure. Consequently, the corresponding bands remain degenerate at the *Γ*-point and exhibiting Dirac-like dispersion along the *x*-direction, shown in the top row of the Fig. 2b. Conversely, the finite interlayer coupling introduces the higher-order scattering involving off-lateral momentum modes ($m \neq 0, n \neq 0$, blue dots). This higher-order scattering generates a quadratic dispersion band emerging from the moiré lattice, as shown in the bottom row of the Fig. 2b. These theoretical predictions are confirmed by RCWA simulations (See Supplemental Materials).

During the inter-dimensional crossover from 1D to 2D driven by the increased interlayer couplings, the hybridization between the Dirac-like bands (originated from the 1D grating) and quadratically dispersive bands (from the moiré lattice) triggers a Lifshitz transition[35]. At the transition point, the change of the Fermi surface topology inverts the band curvature at Γ ($m^* = \partial^2\omega/\partial k^2|_{k=\Gamma}$), which underpins the mechanism for observed flat-band formation.

As the twist angle is varied, lateral momenta remain fixed along the constant-energy contour in the BZ. Meanwhile, the off-lateral momenta undergo an energy shift dependent on the twisted angle due to the changing orientation of the reciprocal lattice vectors, with $\vec{b}_1 \cdot \vec{b}_2 = |\vec{b}_1||\vec{b}_2|\cos(\theta_t)$. Notably, when these shifted quadratic bands energetically resonate with the Dirac bands, the strong band hybridization is pronounced leading to the band flattening. This resonance condition, denoted by $\theta_r = 2\arcsin\left(\frac{p}{2q}\right)$ ($p$ and $q$ are integers) arises purely from geometric considerations and remains insensitive to the specific material parameters.

To theoretically capture these phenomena, we introduce a Bloch Hamiltonian $h(\vec{k})$ expressed within the coupled-mode basis as[36],

$h(\vec{k}) = \sum \left(a(\vec{k}+\vec{g})^2 \delta_{\vec{g},\vec{g}'} + W_{\vec{g},\vec{g}'}\right)$ where $\vec{g} = m\vec{b}_1 + n\vec{b}_2$, $a$ is a mass constant, and $W_{\vec{g},\vec{g}'}$ represents the coupling strength between different BZs. We focus on the leading-order coupling model by considering the coupling between the nearest-neighbor BZs (-1 ≤ $m, n$ ≤ 1), as depicted in Fig. 2a. The effective Hamiltonian takes the explicit form as,

$$h_{eff}(k_x) = \begin{pmatrix} v_{eff}k_x & U & V \\ U & -v_{eff}k_x & V \\ V & V & \frac{1}{2m}k_x^2 + E_{shift} \end{pmatrix} \quad (2)$$

where $v_{eff} = cos(\frac{\theta_t}{2})$, $m = 2qb \sin(\frac{\theta_t}{2})$, $E_{shift} = 2qb \sin(\frac{\theta_t}{2}) - pb$, and $U, V$ represent intralayer and interlayer coupling strength, respectively (See Supplementary Materials S.4). By comparing theoretical predictions and experimental observations, we extract Hamiltonian parameters (See supplementary materials for the detailed analysis). Notably, we find that the interlayer coupling strength, $V \sim 0.5U$, approaches that of the intralayer coupling $U$, directly confirming that our system resides in a strongly coupled regime of the moiré superlattice.

The calculated band structures from Bloch Hamiltonian $h(\vec{k})$ with (solid lines) and without (dashed lines) couplings $U, V$ at twist angles $\theta_t = 46.5°$ and $63.0°$ are shown in Fig. 2c and Fig.2d, respectively. The flat bands do not form exactly at the resonance condition ($E_{shift} = 0$) but instead arise when quadratic dispersions are suppressed by coupling effects, adjustable via $E_{shift}$ or twist angle $\theta_t$. Stronger coupling enhances the suppression of quadratic bands, necessitating a larger $E_{shift}$. A larger $E_{shift}$ causes the band crossing between quadratic and Dirac bands to occur at larger momentum, thereby broadening the flat-band range. Consequently, we can control both the magic angle and the momentum width of the flat band with varying coupling strength.

Figure 2e shows the magic angle, identified by the peak of effective mass at the Γ point as a function of various coupling strengths. Enhanced coupling strengths significantly deform bands, shifting magic angles away from first resonance angle ($\theta_r = 60°$, dashed line). Normalization scale factor $u_{flat}$ indicates the interaction strength fitted to the experiment result. Additionally, stronger interactions extend the angle range where the effective mass remains below a certain threshold $m_{th} = \hbar^2 e$, expanding the flat-band range (gray solid lines in Fig. 2e). Figure 2f depicts the relative flat-band width, defined by the momentum range where the absolute value of the flat-band gradient falls below a specific threshold ($|v_{th}| = 1$), normalized by the width at interaction strength $u_{flat}$. Figure 2f reaffirms the theoretical analysis about the relationship between coupling strength and flat-band range. These theoretical predictions are confirmed by RCWA simulations. In the calculation, we change the thickness of Al$_2$O$_3$ layer between the bilayers to tune the interlayer interaction effectively. As the interlayer distance increases, the two magic angles for flat-band formation approach each other, and the flat-band width gradually decreases (see Extended Data Fig. 4).

**Experimental realization of ultra-wide flat bands in TBG structures**

For experimental demonstration, we fabricate the WS$_2$ TBG using a precise two-step electron-beam lithography (EBL) method[37,38], which enables the creation of multiple twist angles in the same flake, while ensuring uniform thickness and accurate angular control. Figure 3a illustrates the fabrication process of the TBG structure: a 50-nm-thick WS$_2$ flake is first exfoliated onto a quartz substrate, followed by EBL patterning and dry etching. A thin Al$_2$O$_3$ layer is then deposited to protect the bottom structure from etching process. Subsequently, a second WS$_2$ flake is transferred on the bottom layer, and overlay lithography and etching are performed to form the top grating with a well-defined twist angle.

Figure 3b presents optical micrographs of an array of fabricated TBGs. The grating structures in the bottom layer are marked with green squares, and the structures in the top layers are marked with red squares. The overlapping region of the TBG structure appears as a brighter gray contrast in the optical micrographs, indicating successful bilayer formation. Further confirmation is provided by scanning electron microscopy (SEM) images (Fig. 3c), which verify that the upper and lower grating layers are well-aligned and maintain consistent twist angles across the sample.

We measure the angle-resolved transmittance spectra of the TBG structures using a Fourier spectroscopy setup, with the TBG aligned to probe dispersion along the *x*-axis (see Fig.1b). Figure 3d and 3e present the measured angle-resolved transmittance spectra near $\theta_t = 42°$ and $\theta_t = 63°$, respectively. As the twist angle varies, the spectra clearly reveal changes in the curvature and convexity of the photonic bands, with flatband formation observed around 1.24 eV and 1.58 eV, respectively. These flat bands persist over an exceptionally wide angular range, approximately from −50° to +50°, significantly exceeding previously reported values in photonic systems[27,30–34].

The experimental observations are in excellent agreement with the theoretical predictions (Extended Data Fig. 1). In particular, the transition of effective mass sign and the resulting avoided crossings between the Dirac-like and quadratic bands are well reproduced, validating the predicted mechanism of flat-band formation in twisted 1D photonic lattices. In addition, multiple magic-angle sequences are also identified, consistent with predictions from the resonance condition. The accompanying Lifshitz transition reflects the topological protection of the flat dispersions at the magic angles[35].

**Intrinsic optical chirality in twisted bilayer gratings**

The TBG structure breaks both in-plane and out-of-plane mirror symmetries due to the relative rotation and vertical stacking of two twisted grating layers. This symmetry breaking enables the emergence of intrinsic optical chirality. To investigate this effect, we measure the angle-resolved circular dichroism (CD) spectrum. In Fig. 4a, the TBG for $\theta_t = 63°$ exhibits a pronounced CD of about 0.45 at the flatband energy of 1.6 eV, indicating strong optical spin-selective light-matter interaction. A comparative measurement on the 42° TBG sample, which also exhibits optical chirality but with weaker CD, is shown in the Extended data 5.

To verify whether the observed CD at 1.6 eV arises from intrinsic chirality rather than geometric projection effects, we analyze the full transmittance matrix under circularly polarized illumination (Fig. 4b). The matrix consists of four components: $T_{LL}$ and $T_{RR}$, representing spin-preserving transmittance for left and right circular polarization (LCP and RCP) and $T_{LR}$ and $T_{RL}$, representing polarization conversion between LCP and RCP. At the flat band of energy of 1.6 eV, only the $T_{LL}$ component exhibits a sharp transmittance dip, while $T_{RR}$ remains flat, and $T_{RL}$, $T_{LR}$ are nearly zero—indicating that the chiral flatband mode couples selectively and directly to LCP without inducing polarization conversion.

We further validate these features by examining the transmittance matrix at normal incidence, presented in Figs. 4c and 4d, which represent cross-sectional data from Fig. 4b. Both experimental and simulated spectra consistently show the same optical spin-selective signature: a pronounced dip in $T_{LL}$ at 1.6 eV, with all other components remaining featureless. These results confirm that the chiral mode is helicity-preserving and cannot be attributed to asymmetric excitation, supporting its classification as an intrinsically chiral photonic flat band.

To understand the origin of this intrinsic chirality, we perform finite element method (FEM) simulations of the near-field distribution. As shown in Fig. 5e, under LCP illumination, the electric field is strongly confined in the overlapping region of the TBGs, whereas RCP illumination results in relatively negligible field confinement. We quantify the local chirality by calculating optical chirality density (OCD), defined as $\tilde{\chi}_l = (\frac{c}{\omega})Im[D^* \cdot B]$ [39]. The simulated OCD distribution in Fig. 4f reveals intense local angular momentum density localized at the overlapping grating region, consistent with the formation of a spin-polarized eigenmode. This near-field analysis provides compelling evidence that the observed chiral flatband arises from intrinsic optical chirality, rather than extrinsic asymmetries.

**Conclusions**

In summary, we realize the TBG in the strong interlayer coupling regime that is unattainable in electronic systems. The enhancement of the interlayer interaction is attributed to the strong wavefunction overlapping between layers in the photonic platform. Moreover, we emphasize that the flat-band range of the observed flat band is exceptionally wide, even compared to previously reported photonic platforms. This is primarily attributed to the high refractive index of the WS$_2$ layer. When using other optical materials for the same TBG structure—such as hBN[40], a typical vdW material—the flat band width becomes narrower (See Extended Data Fig.6). This behavior can be understood by considering that the slope of the Dirac-like band (i.e., the group velocity) is inversely proportional to the refractive index. A lower index leads to a steeper Dirac-like dispersion and a reduced effective mass in the quadratic band, thereby limiting the extent of the flat band in momentum space. A key advantage of bilayer photonic crystals based on vdW materials is their combination of structural tunability and ease of fabrication, while offering greater flexibility in controlling interlayer distance, twist angle, and lateral alignment—critical parameters for flat-band engineering.

Our theoretical model suggests that flatband formation does not necessarily require a twisted bilayer structure. For example, a single-layer diamond-patterned photonic crystal, which integrates the geometry of two layers into one patterned layer, can exhibit a band structure similar to that of a twisted bilayer system. Our analysis of TBG provides fundamental insights into the conditions that foster flat-band emergence, offering a robust theoretical framework for designing novel geometries that support flat bands. Nonetheless, a bilayer configuration introduces optical chirality, which is absent in the single-layer counterpart, making it particularly advantageous for applications that require chiral optical response.

Furthermore, the ability to manipulate flat bands in photonic systems holds significant potential for applications requiring a high optical density of states, such as enhanced Purcell effects[41,42], lasing[18,34], and other strong light-matter interaction phenomena[18]. Our work not only provides new insights into flat-band engineering but also opens pathways for novel photonic applications leveraging the unique properties of moiré photonic systems.

The flexibility of photonic structures plays a key role in this work. Unlike atomic systems, photonic platforms allow for the design of complex geometries that are otherwise unfeasible. The intrinsic localization associated with the flat-band formation highlights how the interplay between strong interlayer coupling and non-trivial intralayer ordering can stimulate novel research directions within condensed matter physics. Our work thus presents an ideal platform for investigating future moiré-based materials.

## Acknowledgements

M.J.P. thanks to helpful discussions from Chang-Hwan Yi and Hee Chul Park. This work was supported by the National Research Foundation of Korea (NRF) grant (No. RS-2024-00335222 for K.U.) funded by the MSIT of Korea government and Samsung Science and Technology Foundation (SSTF-BA1902-03 for K.U.). This work was supported by the National Research Foundation of Korea (NRF) grant funded by the Korea government (MSIT) (Grants No. RS-2025-16070482, RS-2023-00218998, RS-2025-25446099, RS-2025-03392969 for H.U.).

## Author Contributions

D.C. and S.-J.L. equally contributed to this work. D.C. and S.-H.G. designed and conceptualized this work. D.C., S.-J.L., S.C., D.-J.S. fabricated the sample structures and conducted optical experiments and data collections. D.C., S.-J.L., S.C. conducted numerical simulations. D.C., S.J.L, S.C., C.B.B., S.P. performed all data analysis and visualization. C.B.B., S.P., M.J.P developed the theoretical model. D.C., S.-J.L., S.-H.G. M.J.P. wrote the original manuscript with input from all authors. All authors reviewed and revised the manuscript. M.J.P. and S.-H.G. supervised this project.

## Corresponding authors

Correspondence to Moon Jip Park and Su-Hyun Gong

## Competing interests statement

The authors declare no competing interest

## Figures

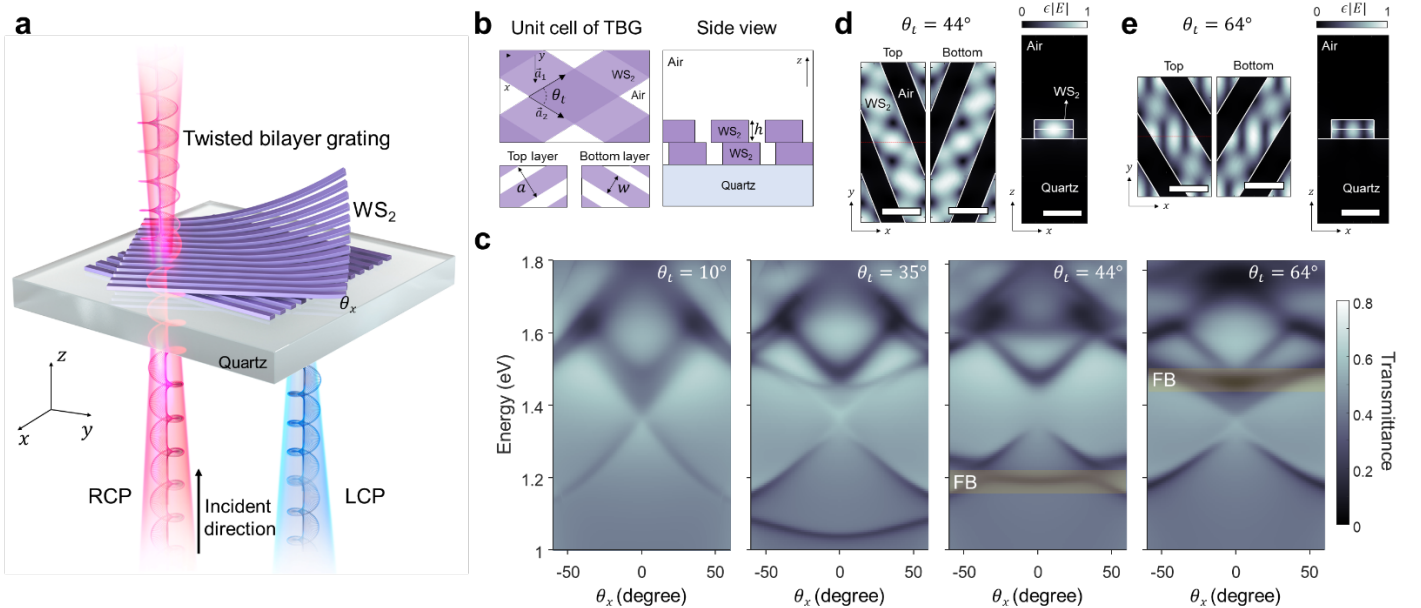

**Fig. 1. Basic schematic of twisted bilayer grating structure. a,** Schematic of TBG structure consisting of two identical WS$_2$ gratings stacked with a twist angle $\theta_t$ between the top and bottom layers on a SiO$_2$ substrate. Incident light propagates along the $z$-direction. **b,** Geometry of a unit cell with top and side views. In the top view, the grating period $a$, grating width $w$, and twist angle $\theta_t$ are defined, with $\vec{a_1}$ and $\vec{a_2}$ indicating the primitive lattice vector of the TBG superlattice. In the side view, each WS$_2$ grating has thickness $h$. **c,** Simulated transmittance spectra of the twisted bilayer grating for various twist angles, at $\theta_t = 10°$, 35°, 44°, and 64° along the $x$-axis. Flat bands (FB) are indicated with yellowish boxes. **d, e,** Simulated electric-displacement distributions for $\theta_t = 44°$ (d) and 64° (e). Left and middle panels show the electric-field profiles in the $x$-$y$ plane taken at the central positions of the top and bottom WS$_2$ layers, while the right panels show the $x$-$z$ cross-sections along the red dashed line. White scale bars represent 200 nm.

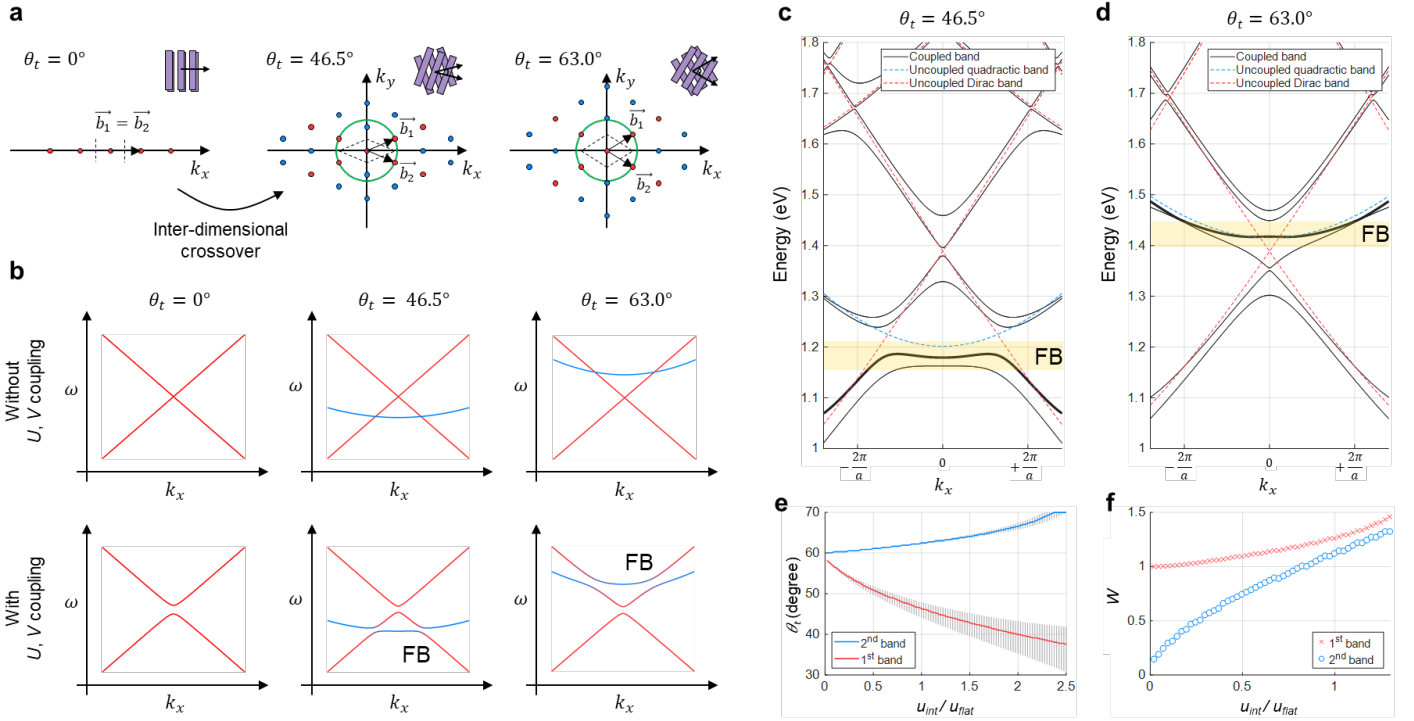

**Fig. 2 Formation of flat band dispersion. a,** Schematic of Brillouin-zone folding in reciprocal space as a function of twist angle $\theta_t$. The reciprocal vectors **b₁** and **b₂** define the Brillouin zone of superlattice. Red (blue) dots indicate the zone centers of the Dirac-like (quadratic) band, while the green circles indicate iso-energy contours. **b,** Illustrative dispersions for different twist angles $\theta_t$, showing Dirac-like (red) and quadratic (blue) bands. The top row shows the noninteracting case, while the bottom row shows the effect of band coupling, which leads to the formation of flat bands (FB). **c, d,** Calculated band structures at twist angles $\theta_t = 46.5°$ and $63.0°$, in the absence (dashed lines) and presence (solid lines) of band coupling. Hybridization between the Dirac and quadratic bands forms a flat band, which is indicated with yellowish boxes. **e,** Magic angle versus the inter-band coupling strength. Colored lines (red and blue) correspond to the flat band in **c** and **d**; the gray region indicates where the effective mass $m_{eff} > 1$. **f,** Relative flat-band width $W$ versus coupling strength for the red and blue bands, matching the color in **e**.

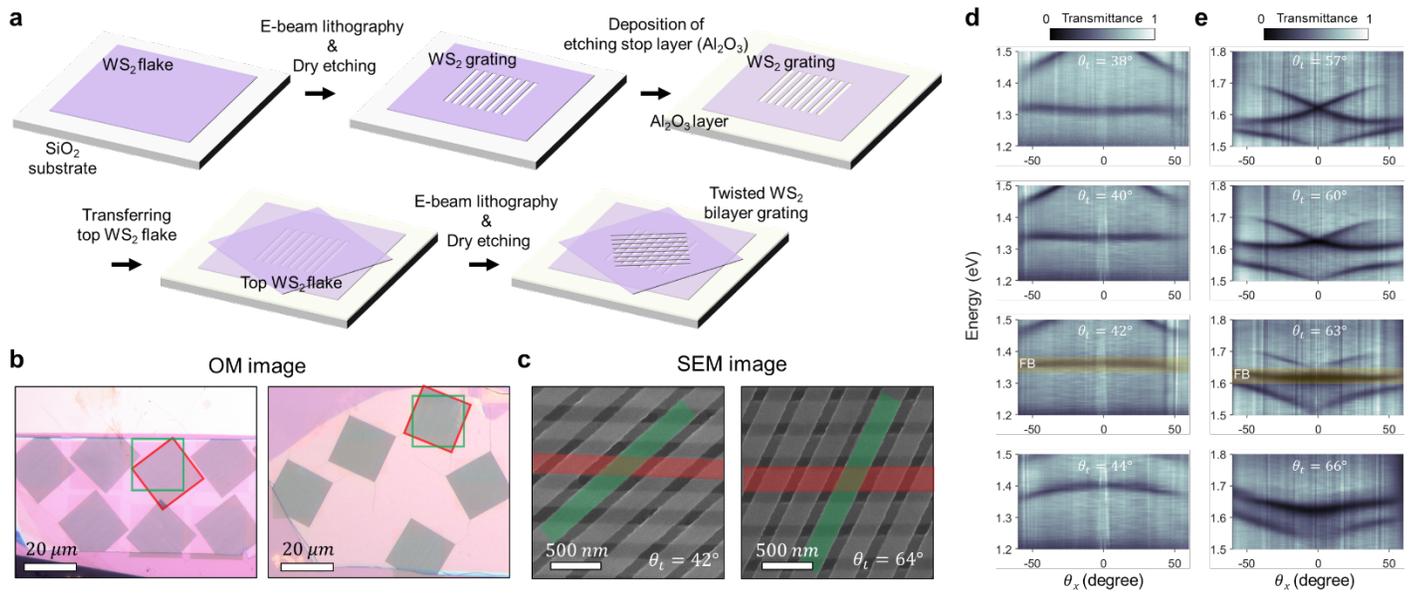

**Fig. 3 Experimental realization of flat band a,** Schematic illustration of the fabrication process for the twisted bilayer grating (TBG) structure. A 50-nm-thick WS₂ flake is exfoliated onto a quartz substrate, patterned into gratings using electron-beam lithography (EBL) and dry etching. A protective Al₂O₃ layer is deposited via atomic layer deposition (ALD), followed by the transfer of another WS₂ flake. A second EBL and etching step is used to define the top grating with a desired twist angle. **b,** Optical microscope (OM) image of the TBG. The blue and red squares indicate the bottom and top grating regions, respectively. **c,** Scanning electron microscope (SEM) image of the TBG structure. **d-e,** Angle-resolved transmittance spectra of the TBG for different twist angles near $\theta_t = 42.0°$ and $63.0°$, respectively.

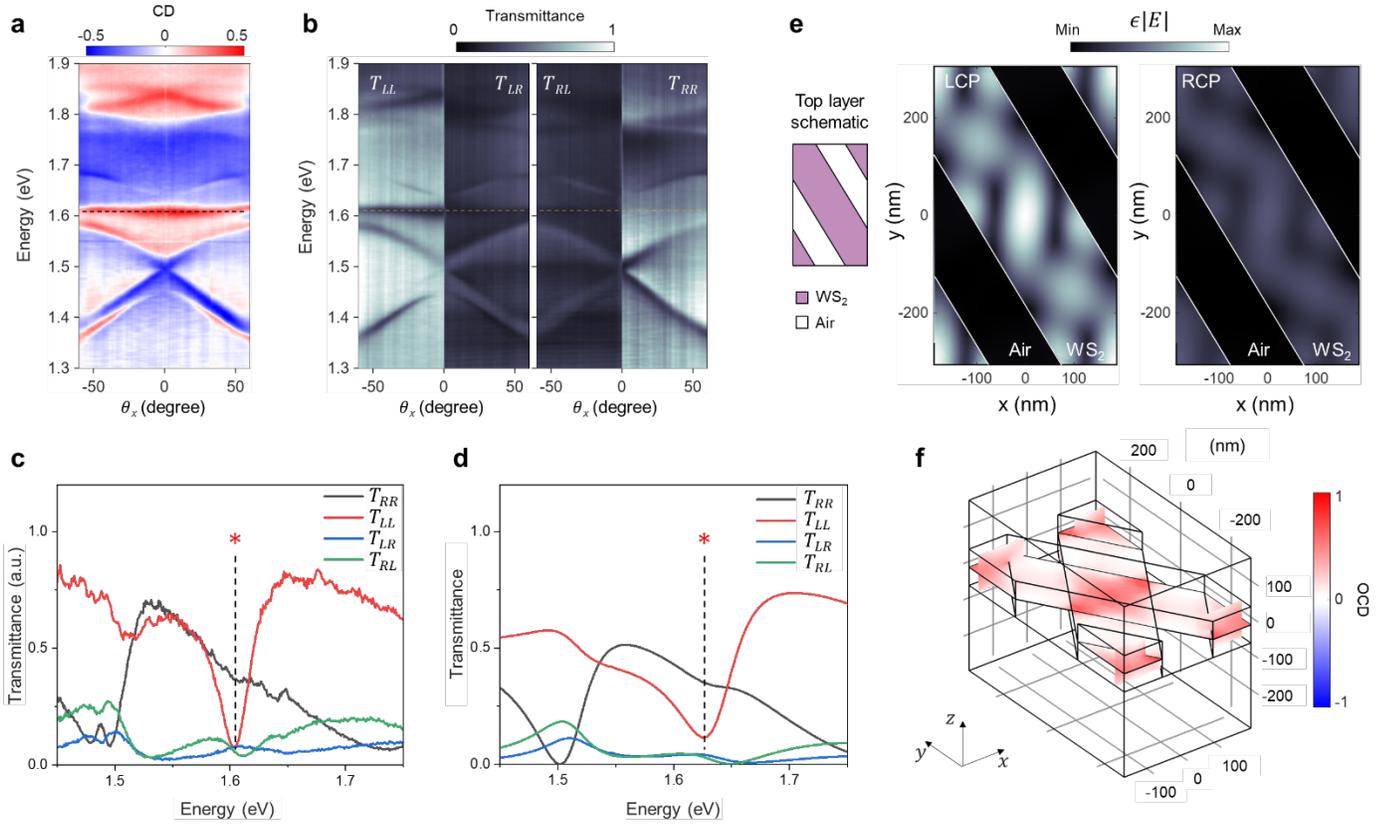

**Fig 4. Intrinsic optical chirality in twisted bilayer gratings. a,** Measured angle-resolved CD dispersion. The dashed line denotes the energy of the chiral flatband mode. The dashed line indicates the chiral flatband resonance at 1.6 eV. **b,** Measured angle-resolved transmittance matrix components ($T_{RR}$, $T_{RL}$, $T_{LR}$, $T_{LL}$) under circularly polarized illumination. Due to mirror symmetry in the momentum space, each component is plotted by combining two symmetric halves. **c,** Polarization-resolved transmittance spectra measured at normal incidence ($k = 0$). The asterisk (*) marks the spectral dip associated with the chiral flatband resonance. **d,** Numerically calculated, polarization-resolved transmittance spectra at normal incidence. The asterisk (*) highlights the resonance position of the chiral flatband at 1.6 eV. **e,** Simulated electric displacement distribution at 1.6 eV under right- and left-handed circularly polarized (RCP and LCP) excitation. The field is strongly localized between the two gratings under LCP illumination, indicating spin-selective mode excitation. **f,** Optical chirality density (OCD) distribution of the chiral eigenmode at 1.6 eV, showing strong chiral field localization within the overlapping region of the two gratings.

**Method**

**Sample Fabrication**

Twisted bilayer grating (TBG) structures were fabricated using 50 nm-thick multilayer $WS_2$ flakes through a series of nanofabrication processes. The $WS_2$ flakes were mechanically exfoliated from bulk crystals and transferred onto a quartz substrate using a polydimethylsiloxane (PDMS) stamping method. To fabricate the bottom grating structure, electron-beam lithography (EBL) was performed at 30 keV on a positive e-beam resist (AR-P 6200.09, ALLRESIST) spin-coated at 4000 rpm for 45 seconds, followed by a soft bake at 180 °C for 1 minute. To suppress surface charging during EBL, a conductive polymer (Espacer 300Z, RESONAC) was spin-coated at 2000 rpm for 40 seconds. The developed patterns were obtained using amyl acetate for 7 minutes, followed by a rinse with isopropyl alcohol (IPA) for 30 seconds. Inductively coupled plasma reactive ion etching (ICP-RIE) was then used to etch the grating structure on the $WS_2$ flakes. The etching was performed under the following conditions: $SF_6$ (40 sccm) and $O_2$ (10 sccm) gas flow, with a bias and source power of 50 W, for a duration of 40 seconds. After etching, the remaining resist was removed using a resist remover (AR 600-71, ALLRESIST). A conformal 5 nm-thick $Al_2O_3$ protective layer was deposited via atomic layer deposition (ALD) to encapsulate and protect the bottom grating structure. To fabricate the top grating, an additional $WS_2$ flake was transferred onto the $Al_2O_3$ layer. Alignment marks enabled angle-controlled EBL to fabricate the top grating at a desired twist angle relative to the bottom layer. A similar etching process was then performed; however, the underlying $Al_2O_3$ layer served as an etch stop, preventing damage to the bottom grating. Finally, any residual resist was removed, completing the fabrication of the twisted bilayer $WS_2$ grating structure.

**RCWA simulation**

Angle-resolved transmittance spectra of the TBG structure were calculated using Rigorous Coupled-Wave Analysis (RCWA) developed in MATLAB. The used parameters are as, grating period $a$ is 320 nm, grating width $w$ is 192 nm, and grating thickness $h$ is 50 nm for each layer. We assume the cover medium is air ($n_{air} = 1$), and the substrate is quartz with a refractive index $n_s = 1.48$ in the calculation. We assume permittivity of $WS_2$ materials using the following Lorentz oscillator model, $\epsilon(E) = \epsilon_\infty + \sum \frac{f_m}{E_m^2 - E^2 - i\gamma_m E}$, where E is an energy, $E_m$ is resonance energy, $f_m$ is oscillator strength, $\gamma_m$ is damping coefficient for m-th oscillator. $\epsilon_\infty = 12$, $E_1 = 1.97$ eV, $E_2 = 2.4$ eV, $E_3 = 2.69$ eV, $E_4 = 2.95$ eV, $f_1 = 0.83$ eV$^2$, $f_2 = 1.38$ eV$^2$, $f_3 = 3.18$ eV$^2$, $f_4 = 30$ eV$^2$, $\gamma_1 = 0.04$ eV, $\gamma_2 = 0.18$ eV, $\gamma_3 = 0.16$ eV, $\gamma_4 = 0.52$ eV[43]. The resultant refractive index of $WS_2$ is plotted in the Extended Data Fig. 7.

**Finite element method analysis**

The near-field distributions and eigenmodes of the TBG and diamond-shaped photonic crystal structures were calculated using three-dimensional finite-element method (FEM) simulations implemented in COMSOL Multiphysics. The TBG structures were modeled as laterally infinite arrays on a quartz substrate. Bloch periodic boundary conditions were applied along the x- and y-directions, while perfectly matched layers (PMLs) were used along the *z*-direction to prevent spurious reflections. To identify chiral eigenmodes, the eigenfrequency solver was employed to extract their resonant frequencies and associated quality factors. To simulate the circular dichroism observed in angle-resolved transmittance experiments, a circularly polarized plane wave was incident from the substrate side at a tilted angle. The transmitted power, represented by the Poynting vector, was integrated over the simulation domain to obtain the angle-resolved transmittance spectrum.

**Measurements**

The angle-resolved transmittance spectrum is measured using a broadband plasma light source (XWS-30, ISTEQ) and the Fourier optics technique (see Extended Data Fig. 8). We utilize two microscope objective lenses to focus white light on the structure and to collect the transmitted light. The objective lens for focusing the white light has a magnification of 20x and a numerical aperture (NA) of 0.95, and the objective lens for collecting the transmitted light has a magnification of 40x and an NA of 0.95. The transmitted light collected by the objective lens is collimated through a 4F lens alignment for obtaining a Fourier image and the signal was measured by a spectrometer with a charge-coupled device detector. And we use the vertical entrance slit of the spectrometer to slice only the $\theta_y = 0$ of the signals. Moreover, a quarter-wave plate, a half-wave plate and a linear polarizer are used to select the polarization of light.

**Data availability**
The datasets used in this study are available

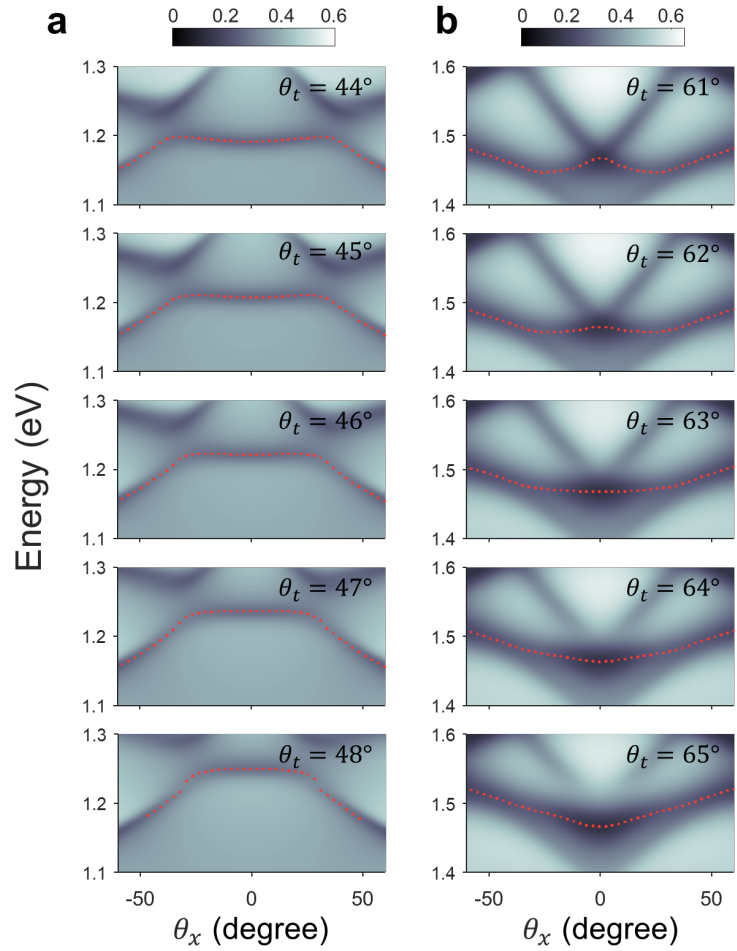

**Extended Data Fig.1 Evolution of flat-band dispersions. a** and **b,** Calculated transmittance spectrum of TBG around $\theta_t = 46.0°$ and $\theta_t = 63.0°$, respectively. Red dots mark the band dispersions, and as the twist angle increases, the curvature of the bands evolves.

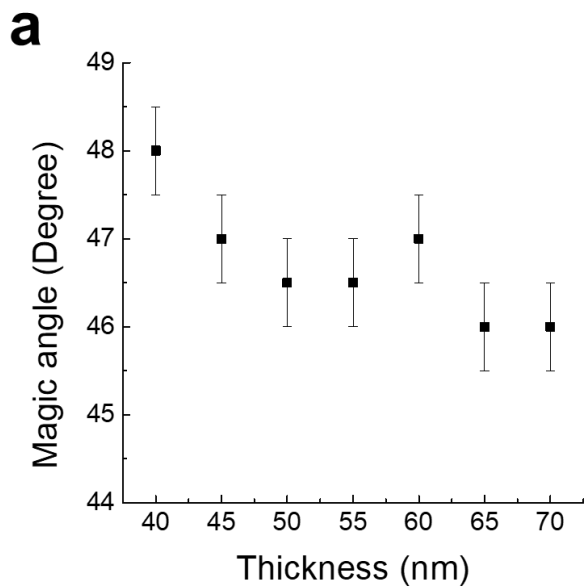 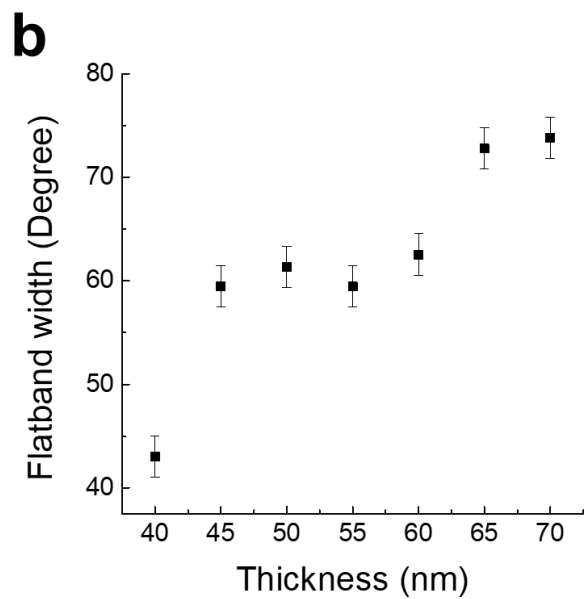

**Extended Data Fig.2 Flat-band characteristics as a function of thickness of TBG** Thickness dependence of (a) magic angle and (b) flat-band range in TBG.

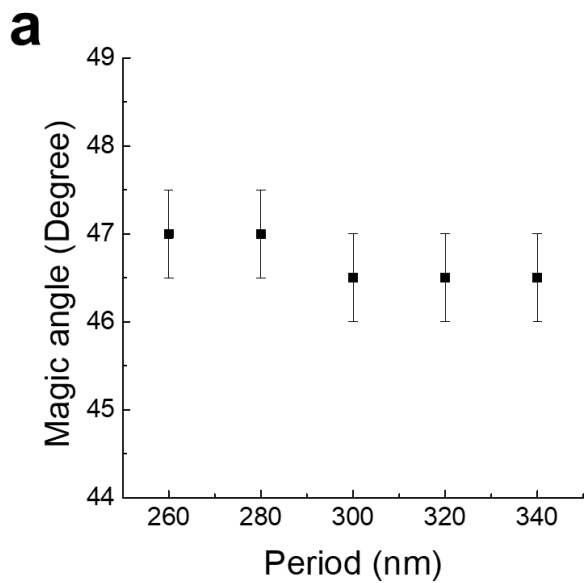 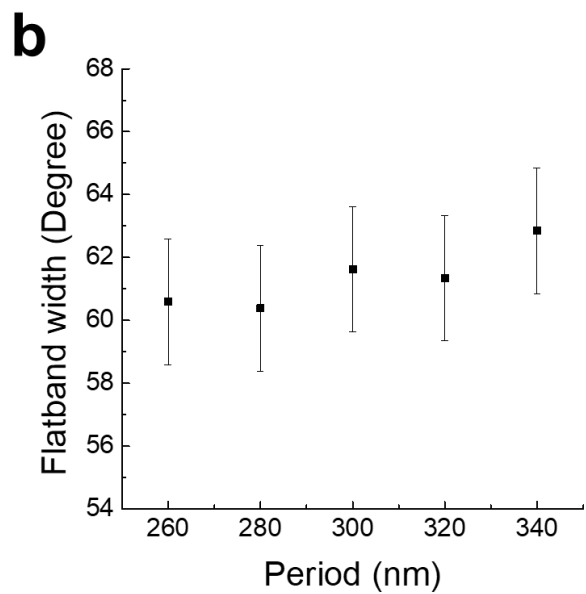

**Extended Data Fig.3 Flat-band characteristics as a function of grating period of TBG** Grating period dependence of (**a**) magic angle and (**b**) flat-band range in TBG.

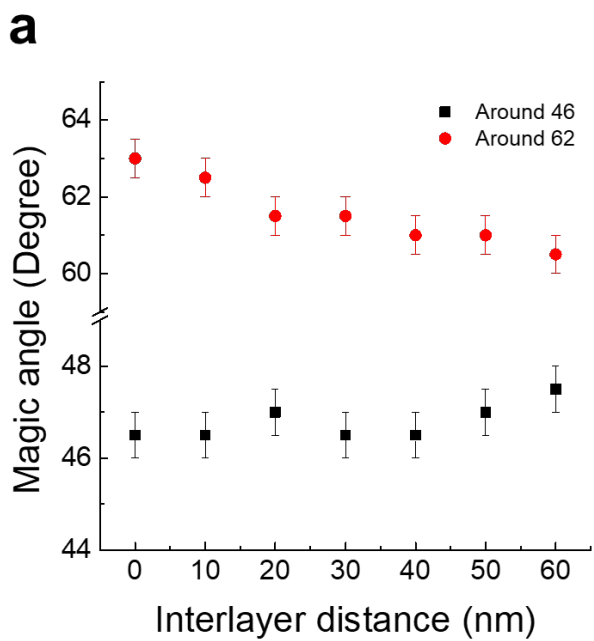 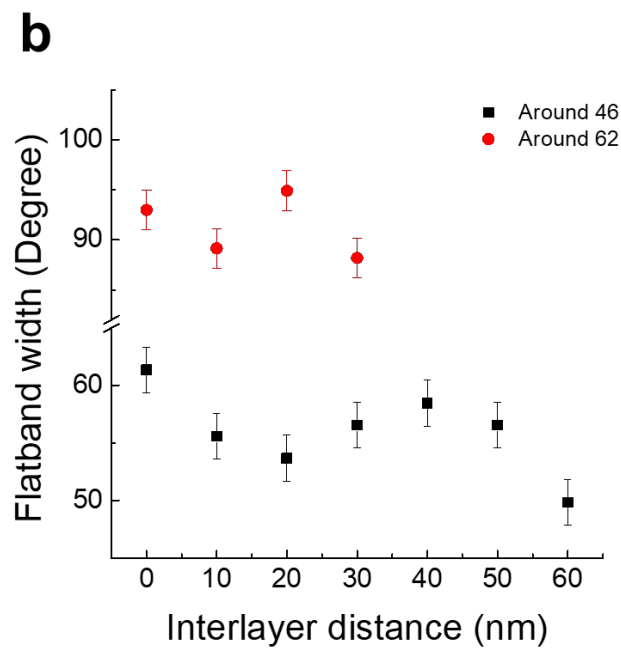

**Extended Data Fig.4 Evolution of flat-band dispersions.** Calculated transmittance spectrum of TBG around $\theta_t = 46.0°$ and $\theta_t = 63.0°$, respectively. Red dots mark the band dispersions, and as the twist angle increases, the curvature of the bands evolves.

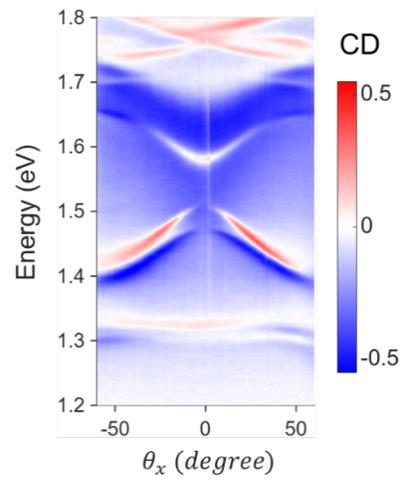

**Extended Data Fig.5 Angle-resolved circular dichroism of TBG for 42°**

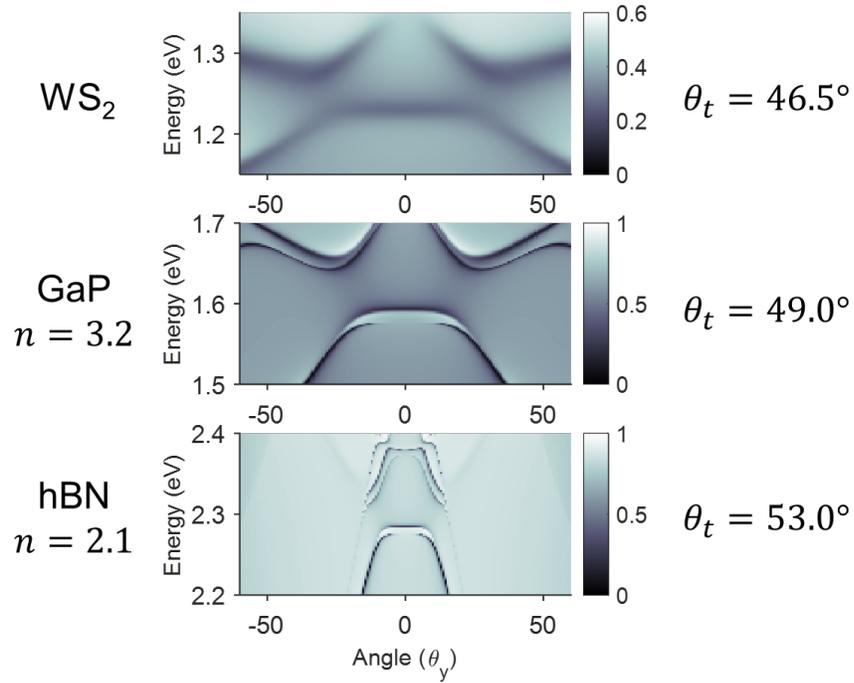

**Extended Data Fig.6 Flat-band behavior of TBG using WS₂, GaP, and hBN.** Numerical simulations with different refractive indices to investigate the effect of refractive index (material) in the TBG structure. For comparison, we select two more materials with lower refractive indices: GaP (n ≈ 3.2) and hBN (n ≈ 2.1). The resulting transmittance spectrum reveals that the flat-band width is highly sensitive to the refractive index of the constituent material.

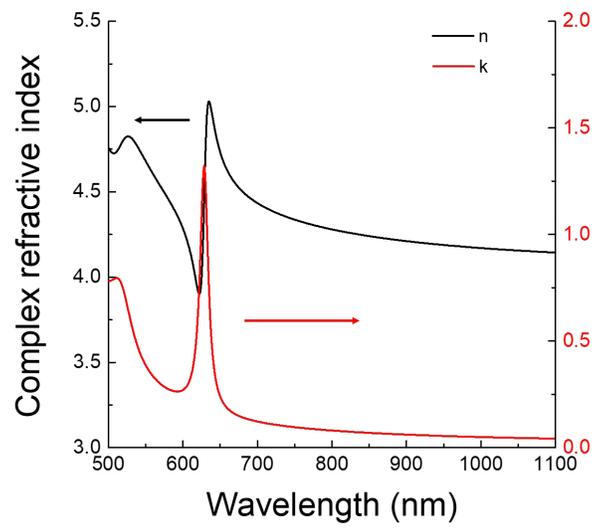

**Extended Data Fig.7 Refractive index of WS$_2$ for numerical calculation**

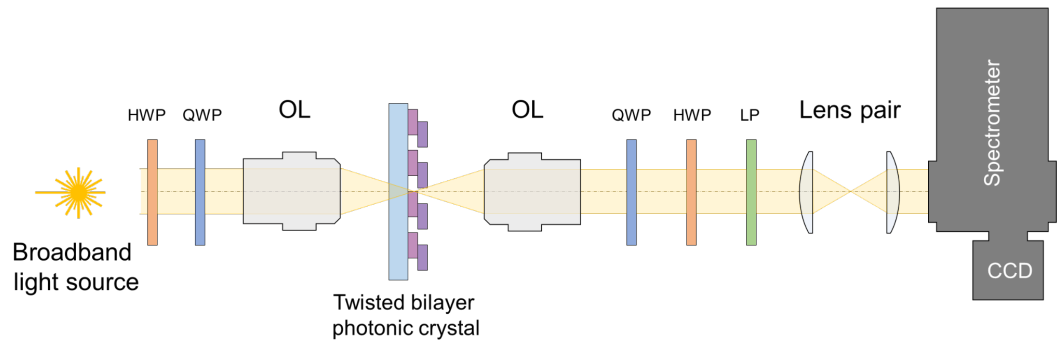

**Extended Data Fig.8 Schematic of the angle-resolved transmittance spectroscopy setup.** A broadband white light source is employed to probe the transmittance spectrum through the sample. The optical path includes two microscope objective lenses (OL), and a pair of lenses for angular dispersion. Polarization of light is manipulated with half-wave plates (HWP), quarter-wave plates (QWP), and a linear polarizer (LP).

# Strong interlayer coupling and chiral flat-band cascades in twisted bilayer gratings

Daegwang Choi[1,†], Soon-Jae Lee[1,†], Seung-Woon Cho[1], Chan Bin Bark[2], Seik Pak[2], Dong-Jin Shin[1], Moon Jip Park[2,*] and Su-Hyun Gong[1,*]

[1]Department of Physics, Korea University, Seoul, 02841, Republic of Korea

[2]Department of Physics, Hanyang University, Seoul, 133-791, Korea

## S.1 Evolution of BZ and shift of quadratic bands

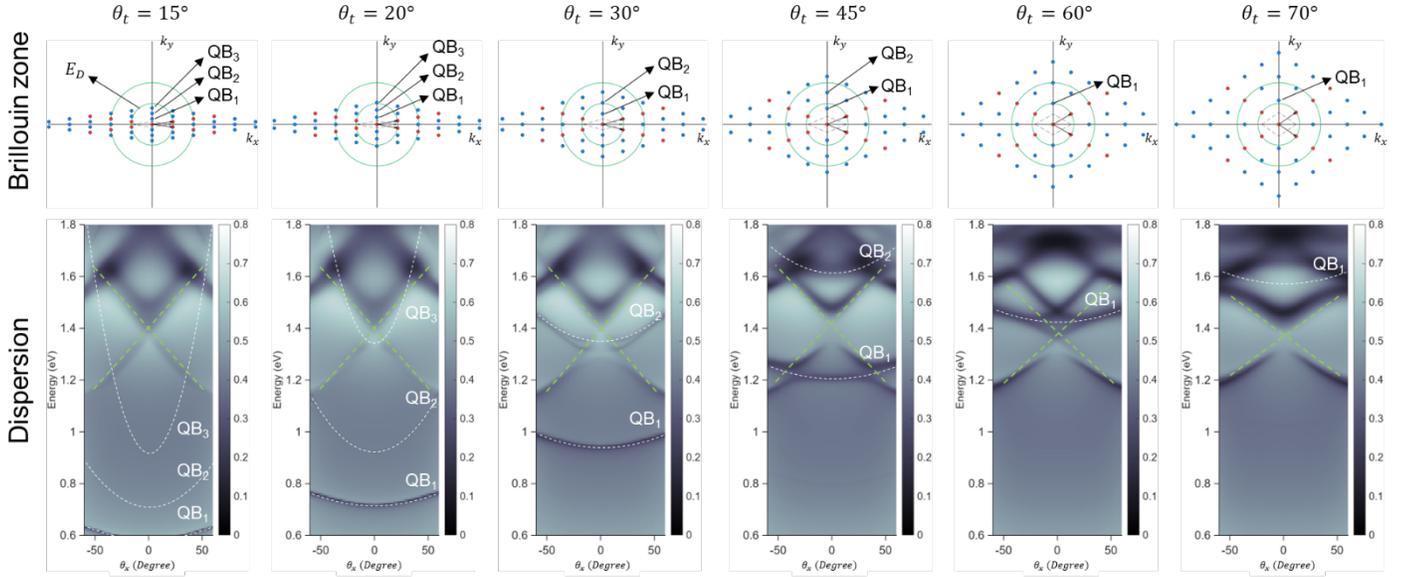

**Fig.S.1**. Evolution of quadratic bands (QB) as a function of twist angle $\theta_t$. Top row: reciprocal lattice diagrams showing the positions of quadratic band points (QB$_{1,2,3}$) relative to the iso-frequency circle (green) for different $\theta_t$. Bottom row: simulated angle-resolved spectra with guidelines (white dashed lines). When a quadratic band point lies near the iso-frequency circle, it couples to the Dirac band. As $\theta_t$ increases, higher-order quadratic bands (QB$_3$ and QB$_2$) shift away from the Dirac band and disappear. In the intermediate range of $\theta_t$, quadratic bands intersect with the linear Dirac-like guided bands (green dashed lines), leading to strong interaction and hybridization. This accounts for the transition from multiple quadratic bands to a single quadratic band (QB$_1$) at certain twist angle $\theta_t$.

## S.2 How to define a magic angle in a TBG structure?

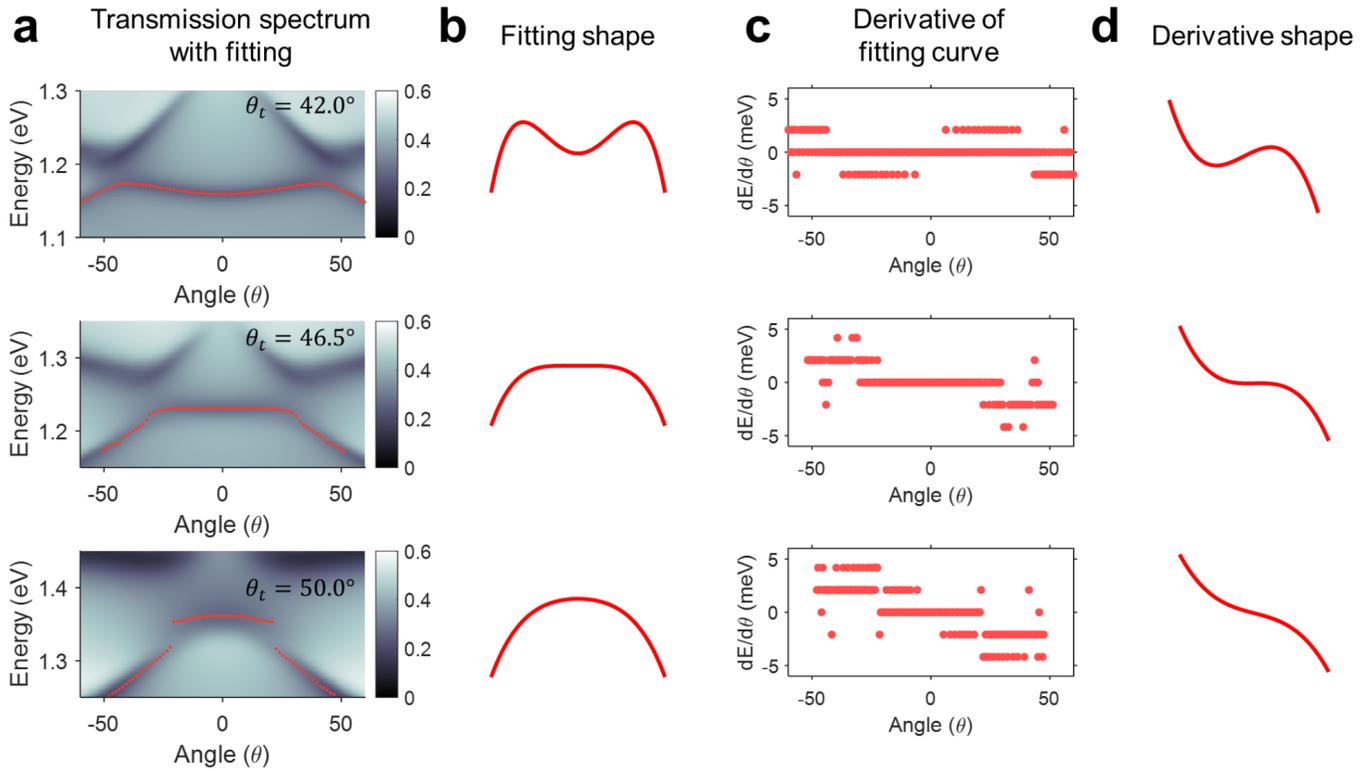

**Fig.S.2.** Method to define a magic angle in a TBG structure. (a) Transmittance spectra depending on the twist angle with fitting. (b) Schematic of fitting curve shape (c) Derivative of fitting curve depending on the twist angle. (d) Schematic of derivative data shape.

We define the magic angle in the TBG structure as the twist angle at which the photonic flat band becomes maximally flat in the numerical calculation. The magic angle corresponds to the point where the curvature of the convex photonic dispersion inverts. Figure S.2 illustrates this definition with representative examples. Figure S.2 (a) shows the transmittance spectra for three twist angles: 42.0°, 46.5°, and 50.0°. The dip points in the photonic dispersion are indicated by red dots. A schematic sketch of the dispersion is presented in Figure S.2 (b). At 42.0°, the dispersion exhibits an M-shaped (or W-shaped) profile, which gradually evolves into an upward-convex shape as the twist angle increases. To quantify this transition, we compute the derivative of the fitted dispersion curves in the transmittance spectra. The derivative profiles are shown in Figure S.2 (c). We define the magic angle as the twist angle at which the sign of the second derivative at $\theta = 0$ changes. Figure S.2 (d) provides representative examples of the derivative curves corresponding to the sketched dispersions in Figure S.2 (b). The RCWA calculations were performed with a resolution of 0.5°, so the identified magic angle is determined with an uncertainty of ±0.5°.

## S.3 How to define the width of a flat band?

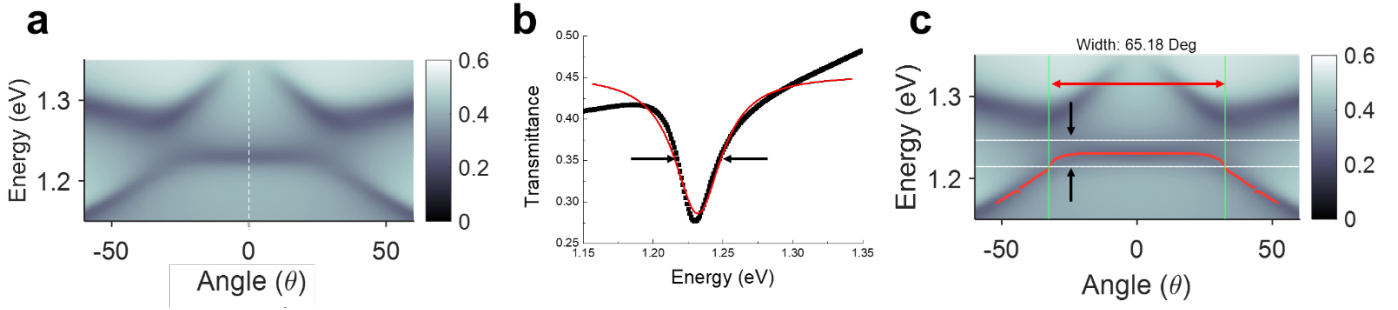

**Fig.S.3.** Method to define the width of flat band in a TBG structure.

We define the width of a flat band numerically in the following way. At the magic angle, we plot the transmittance spectrum at $\theta = 0$, indicated by a white dotted line in Fig. S.3 (a). We then fit the resonance mode using an asymmetric peak fitting function as follows,

$$y = \frac{A\left(q + \frac{x - x_c}{w}\right)^2}{1 + \left(\frac{x - x_c}{w}\right)^2} + B$$

Where $A$ represents the amplitude, $B$ is the constant baseline, $x_c$ is the resonance center position, $q$ is the asymmetry factor, and $w$ is the linewidth. Based on this, we define the flat-band energy range, which is marked by white dotted lines in Fig. S.3 (c). The flat-band width is then determined as the energy span between the two crossing points within this range, indicated by light green lines in Fig. S.3 (c).

As this method may overestimate the flat-band width for bands with larger linewidths, it is employed only for comparative purposes in this work.

## S.4 Tight-binding model of TBG in momentum space

### Tight-binding model in momentum space

The formation of ultra-wide flat bands in the twisted bilayer grating (TBG) is captured by a tight-binding model in momentum space. The periodicity of the moiré superlattice, formed by twisting two one-dimensional gratings by an angle $\theta_t$, is defined by the primitive reciprocal lattice vectors:

$$b_1 = \frac{2\pi}{d}\left(\sin\frac{\theta_t}{2}, \cos\frac{\theta_t}{2}\right) \quad b_2 = \frac{2\pi}{d}\left(\sin\frac{\theta_t}{2}, -\cos\frac{\theta_t}{2}\right)$$

where $d$ is the period of the individual gratings. The system's behavior is described by a Bloch Hamiltonian, $H(k)$, which acts on a vector of field amplitudes, $\Psi$. The eigenvalues of this Hamiltonian correspond to the squared eigenfrequencies of the photonic modes:

$$H(k)\Psi = \left(\frac{\omega(k)}{c}\right)^2 \Psi$$

The basis states for the Hamiltonian are plane-waves $|k + g\rangle$, where $k$ is the crystal momentum within the first moiré Brillouin zone (BZ) and $g = mb_1 + nb_2$ (with $m, n \in \mathbb{Z}$) are the moiré reciprocal lattice vectors. The Hamiltonian is composed of a kinetic term and an interaction term:

$$H(k) = H_0(k) + H_{int}$$

### Kinetic term

The kinetic term is diagonal in the momentum-space basis and describes the free-particle dispersion folded into the moiré BZ. For each momentum state $|k + g\rangle$, its contribution to the eigenvalue is:

$$\langle k + g|H_0(k)|k + g\rangle = |k + g|^2.$$

### Interaction term

The interaction term, $H_{int}$, describes the Umklapp scattering processes induced by the periodic dielectric function of the gratings. The coupling strength between two states $|k + g\rangle$ and $|k + g'\rangle$ depends on the momentum difference $\Delta g = g - g'$. We decompose this interaction into intralayer coupling, $U$, arising from a single grating, and interlayer coupling, $V$, from the interaction between the two gratings. The intralayer term primarily couples momentum states collinearly (e.g., scattering with $\Delta m = 0$ or $\Delta n = 0$ where $\Delta m = |m - m'|$ and $\Delta n = |n - n'|$.), while the interlayer term enables coupling to "off-lateral" modes (where both $\Delta m \neq 0$ and $\Delta n \neq 0$), which is crucial for band flattening.

### Numerical calculation

The band structure calculations were performed by numerically diagonalizing the Hamiltonian $H(k)$. The basis set was constructed using reciprocal lattice vectors $g$.

We define the basis states by the reciprocal lattice vector index pair $(m, n)$ and a two-component layer index. The Hamiltonian is a large matrix composed of $2 \times 2$ blocks, $H_{(m,n),(m',n')}$, which couple the state at reciprocal lattice site $g' = m'b_1 + n'b_2$ to the site $g = mb_1 + nb_2$.

The diagonal blocks, where $(m, n) = (m', n')$, are given by the kinetic term:

$$H_{(m,n),(m,n)} = |k + mb_1 + nb_2|^2 \cdot I_{2\times 2}$$

The off-diagonal blocks, where $(m,n) \neq (m',n')$, represent the interaction potential $W_{(m,n),(m',n')}$. The interaction potential is a sum of intralayer and interlayer terms, $W = W_{intra} + W_{inter}$. The intralayer term is anisotropic in the layer basis:

$$W_{intra} = \delta_{\Delta n,0} U(\Delta m) \begin{pmatrix} 1 & 0 \\ 0 & 0 \end{pmatrix} + \delta_{\Delta m,0} U(\Delta n) \begin{pmatrix} 0 & 0 \\ 0 & 1 \end{pmatrix}$$

where the non-zero intralayer coupling strengths are $U(1) = 1$ and $U(2) = 4$ in the units of the model. The interlayer term is isotropic in the layer basis:

$$W_{inter} = V_{\Delta m, \Delta n} \cdot I_{2\times 2}$$

where the non-zero interlayer coupling strengths $V_{\Delta m,\Delta n}$ are given by: $V_{1,0} = V_{0,1} = 1.25$; $V_{2,0} = V_{0,2} = -0.25$; $V_{1,1} = 0.22$; and $V_{2,1} = V_{1,2} = -0.25$. We then define the $u_{flat}$ as a normalized parameter that represent the current configuration of $W_{(m,n),(m',n')}$.

The full Hamiltonian block is the sum of these kinetic and interaction parts:

$$H_{(m,n),(m',n')} = \delta_{m,m'}\delta_{n,n'}(|k + mb_1 + nb_2|^2)I_{2\times 2} + W_{(m,n),(m',n')}$$

(1)

**Minimal effective Hamiltonian**

The essential physics of band flattening can be captured by a minimal effective Hamiltonian. This model simplifies the full problem by considering only the interaction between a $p$-th order Dirac-like band and a $q$-th order quadratic moiré band. The basis for this 3x3 model consists of states representing these bands. The $p$-th order Dirac-like bands are associated with the nearly-degenerate, linearly dispersing states at momentum points along the principal axes, such as $g = pb_1$ and $g = pb_2$. The $q$-th order moiré band is associated with a quadratically dispersing state at an off-lateral momentum point, such as $g = qb_1 + qb_2$.

We can expand the effective Hamiltonian near the $\Gamma$ point in the form of describing the coupling between these three states as:

$$h_{eff}(k_x) = \begin{pmatrix} v_{eff}k_x & U_{eff} & V_{eff} \\ U_{eff} & -v_{eff}k_x & V_{eff} \\ V_{eff} & V_{eff} & \frac{1}{2m}k_x^2 + E_{shift} \end{pmatrix}$$

Here, the parameters are derived from the full Hamiltonian Eq.S.4(1). The group velocity of the unperturbed Dirac bands is $v_{eff} = \cos(\theta_t/2)$, and the effective mass of the quadratic band is $m = 2qb\sin(\theta_t/2)$, where $b = |b_1| = |b_2|$ is the magnitude of the primitive reciprocal lattice vectors. $U_{eff}$ is the effective intralayer coupling that lifts the degeneracy of the crossing bands, and $V_{eff}$ is the effective interlayer coupling between the Dirac-like band and the quadratic band. The term $E_{shift}$ represents the eigenvalue difference between the p-th Dirac point and the minimum of the q-th quadratic band, which is highly sensitive to the twist angle $\theta_t$.

The magic angle condition occurs when $E_{shift} = 0$, bringing the bands into resonance. The hybridization induced by $V_{eff}$ then causes the flattening of the bands over a wide momentum range.

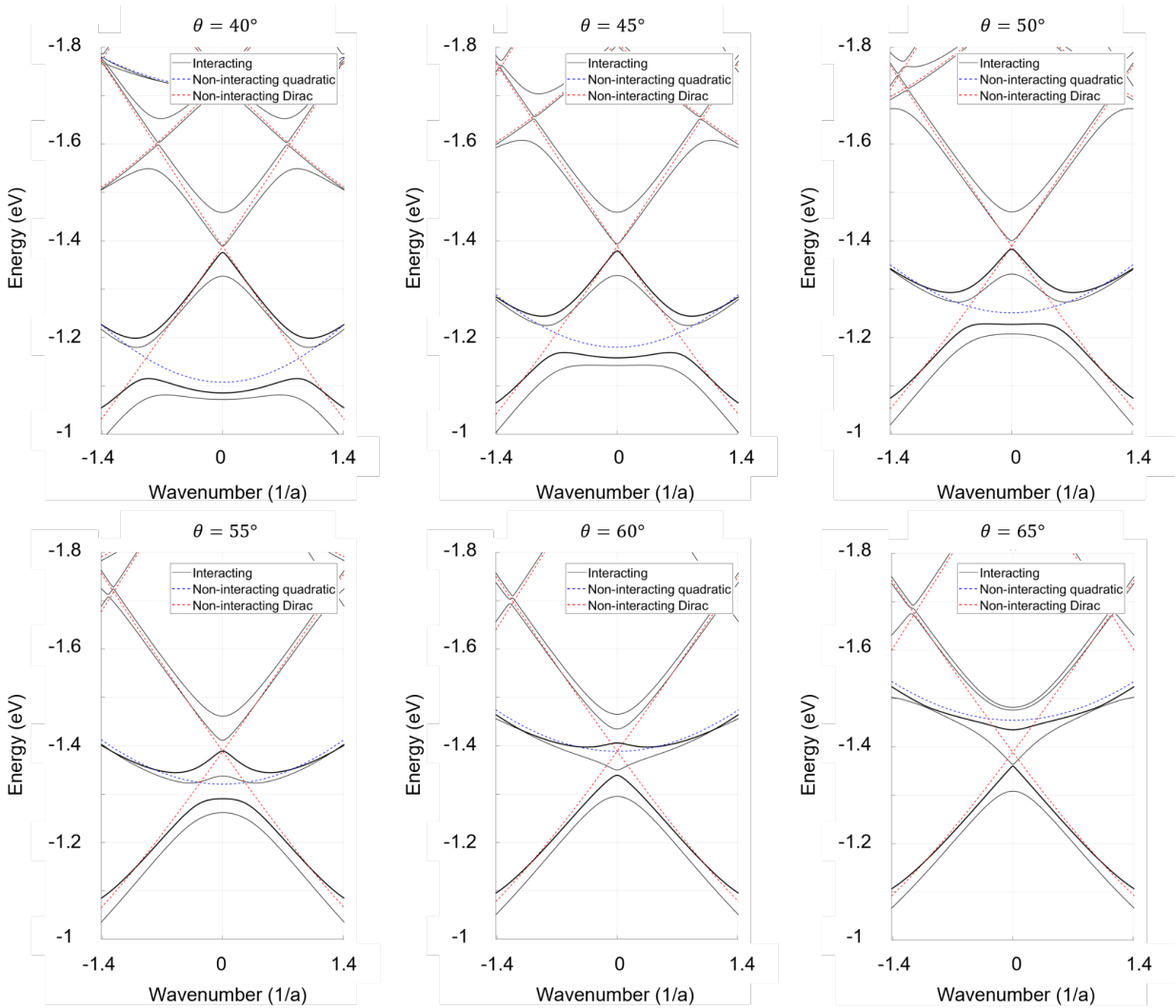

**Fig.S.4. Band structures for various angles from tight-binding model.** Band structure of TBG with respect to the wavenumber (momentum) from the Hamiltonian Eq.S.4(1) in various twist angle $\theta$.

## S.5 Effect of interaction strengths on magic angle and flat-band width

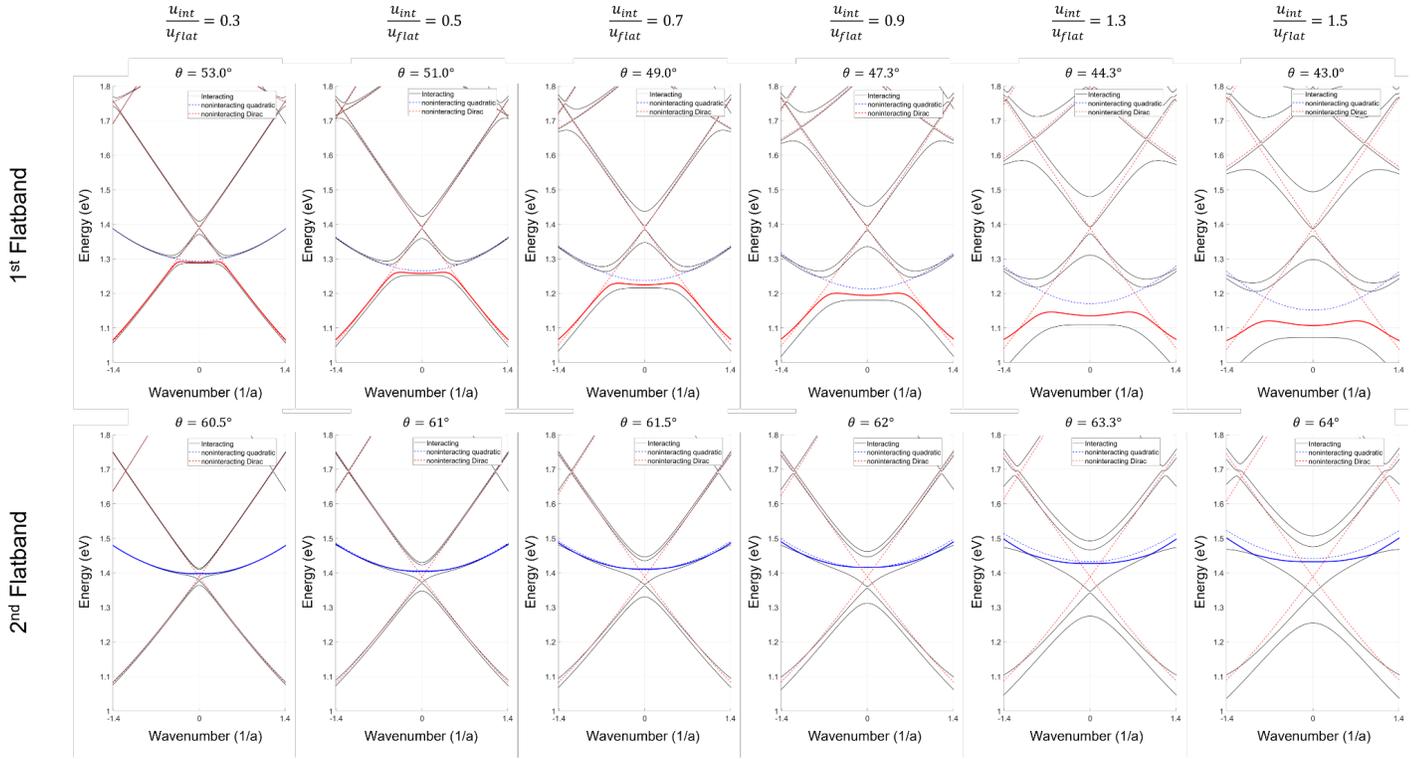

**Fig.S.5. Flat bands for various interaction strengths** Band structure of TBG at magic angles for various normalized interaction strengths $u_{int}$. $u_{int}$ is determined by a uniform scaling factor applied to a baseline set of parameters fitted to the experiment, $u_{flat}$. Thus $u_{int}/u_{flat}$ corresponds to overall change of interaction strength from the experiment fitted parameter. Both 1st and 2nd flat bands show the flat bands broaden as the interaction strength increases.